\documentclass{Odyssey2026}

\makeatletter

\renewcommand{\affiliation}[3]{%
    \if\relax\detokenize{#1}\relax
        \edef\affiliationslist{\affiliationslist\affiliationsep#2, #3}
    \else
        \edef\affiliationslist{\affiliationslist\affiliationsep#1, #2, #3}
    \fi
    \renewcommand{\affiliationsep}{\vskip1pt}
}

\makeatother

\interspeechcameraready

\title{Latent Secret Spin: Keyed Orthogonal Rotations for\\ Blind Speech Watermarking in Anisotropic Latent Spaces}

\author[]{Emma}{Coletta}
\author[]{Massimiliano}{Todisco}
\author[]{Michele}{Panariello}
\author[]{Antonio}{Faonio}
\author[]{Nicholas}{Evans}

\affiliation{}{EURECOM}{France}

\email{[firstname.lastname]@eurecom.fr}


\usepackage{comment}
\usepackage{soul}
\usepackage{tabularx}
\usepackage{graphicx}
\usepackage{subcaption}
\usepackage{longtable} 
\usepackage{colortbl}
\usepackage{multirow, multicol, array, makecell}
\usepackage[advantage,asymptotics,adversary,sets,keys,ff,lambda,primitives,events,operators,probability,logic,mm,complexity,landau,oracles]{cryptocode}
\usepackage{algorithm}
\usepackage[indLines]{algpseudocodex}
\usepackage[hang,flushmargin]{footmisc}
\usepackage[margin=1in]{geometry}
\usepackage{tikz}
\usetikzlibrary{decorations.pathreplacing}
\usepackage{caption}
\usepackage{enumitem}

\begin{document}

\maketitle

\begingroup
\makeatletter
\makeatother



\begin{abstract}

We introduce Latent Secret Spin (LSS), a blind speech watermarking method based on geometric operations in codec latent space. Based upon orthogonal rotations to principal components, LSS induces imperceptible but detectable covariance signatures according to a pseudo-random watermarking schedule. The scheme generalises across datasets, preserves perceptual quality and, unlike some learned, neural watermarking schemes, it does not require neural network training, is resistant to common signal manipulations and is flexible to payload size. Analyses show that structured latent-space watermarking is a promising and interpretable alternative to existing approaches.
\end{abstract}

\vspace{.2cm}

\section{Introduction}
\label{sec:intro}
Astonishing advances in generative artificial intelligence have revolutionised multimedia content creation, but have also increased the risk of misuse, including the spread of misinformation, identity fraud, deepfakes and malicious content manipulation. 
Digital watermarking~\cite{cao2025watermarking} offers a proactive solution and can be used to embed imperceptible yet traceable markers within digital media, enabling the identification of AI-generated content~\cite{cao2025watermarking, charfeddine2022audio}.

In the audio domain, speech watermarking is used to embed imperceptible yet reliably detectable payloads directly into the audio waveform, enabling traceability and integrity verification~\cite{faundez2010speech}.
By incorporating inaudible markers in the audio, watermarking provides a reliable mechanism for verifying the provenance of a speech recording.

Effective, practical speech watermarking strategies must meet three key criteria~\cite{cao2025watermarking, hua2016twenty, lie2006robust}:

\begin{itemize}
    \item \textbf{Imperceptibility:} watermarks must not degrade perceptual audio quality or introduce artifacts that are detectable by a human listener. 
    \item \textbf{Robustness:} they must remain detectable under a range of 
    signal transformations, such as compression, additive noise and filtering. 
    \item \textbf{Security:} they must resist unauthorised removal, spoofing, or forgery under adversarial conditions. 
\end{itemize}
\vspace{.3cm}
Many practical deployment scenarios also require \emph{blind} watermarking. 
A watermarking scheme is considered blind if detection can be performed without access to the original, unwatermarked asset~\cite{hua2016twenty, barni2004watermarking}. 
Blind detection is particularly advantageous and often necessary when the original host asset is inaccessible or unavailable.
Non-blind approaches are then impractical~\cite{charfeddine2022audio}. 

\vspace{.7cm}

We introduce Latent Secret Spin (LSS), a novel principal component-based framework for blind neural speech watermarking.
LSS operates directly in the continuous latent space of a neural audio codec, enabling imperceptible watermark embedding while ensuring robust detection under realistic signal transformations.
The latent space provides a structured representation in which statistical watermark signatures can be introduced and reliably detected~\cite{liu2023dear, wen2025watermark, li25g_interspeech}. Recent approaches, such as WavMark~\cite{chen2023wavmark} and AudioSeal~\cite{roman2024proactive}, demonstrate the effectiveness of learned embedding-detection pipelines for robust and imperceptible speech watermarking, motivating our use of similar structured latent representations.

LSS projects latent representations into a principal component space to exploit its natural geometric structure. 
Watermarks are embedded through small, localised orthogonal rotations within targeted principal component planes.
The embedding process is controlled by a schedule which specifies a set of embedding parameters 
deterministically derived from a secret key, including
the selection of principal component planes, the order in which rotations are applied, and the corresponding rotation angles.
Rotations are distributed across dimensions and time, inducing small but consistent statistical dependencies among principal components that can be exploited for detection. 
This design enables reliable blind watermark detection which is robust to common signal transformations
and unauthorized detection
while preserving perceptual audio quality.

LSS is also lightweight in that it operates on the geometry of pre-trained speech representations and does not require the training of any embedding model.
To the best of our knowledge, LSS is the first speech watermarking method to induce controlled covariance patterns in a principal component space through geometric transformations, and the first to use covariance patterns explicitly for payload embedding.

The paper is structured as follows. 
In Section~\ref{sec:related_work} we provide a review of related work in the literature. 
In Section~\ref{sec:principle} we introduce the geometric principles underlying the LSS framework, while the algorithm itself is described in Section~\ref{sec:implementation}. 
We outline our experimental setup in Section~\ref{sec:setup} and report results in Section~\ref{sec:results}. We discuss the main findings, ongoing research directions, and future work 
in Sections~\ref{sec:discussion} and~\ref{sec:conclusion}.

\section{Related Work}
\label{sec:related_work}

Early digital watermarking techniques operated directly in the signal domains, e.g.\ spatial for images, temporal for audio. 
Alternatively, they
embedded payloads in transform domains using standard representations like the Discrete Cosine Transform (DCT), Discrete Fourier Transform (DFT), or Discrete Wavelet Transform (DWT)~\cite{cao2025watermarking, charfeddine2022audio, hua2016twenty, lie2006robust, celik2005pitch, murata2011audio, dhar2013dwt}. 
Principal Component Analysis (PCA) was used in watermarking 
to remove linear dependencies and decorrelate data~\cite{hien2003pca, chawla2018ARO}. 
PCA was typically paired with frequency-domain transformations such as the DCT or DWT to embed payloads into selected frequency components~\cite{chawla2018ARO,sinha2011digital, saboori2014ANM, Tonge2014ASO}. 
Within all these frameworks, watermarks were generally embedded by modifying dominant principal components or their associated coefficients~\cite{saboori2014ANM, wang2000watermarking, hien2003pca}. Recent works have also explored PCA for feature extraction in zero-shot watermarking schemes~\cite{kahdim2023principal, yang2025zero}. 
In the speech domain, PCA has been used to improve watermark robustness and security by 
separating core speech content from noise and interference, enabling watermark embedding in stable features, such as the formant structure, that are less affected by common signal processing distortions~\cite{wang2018speech, wang2020secure}.

Recent advances in deep learning have led to  neural approaches which encode watermarks into latent representations of speech content, improving both imperceptibility and robustness. 
End-to-end architectures such as WavMark~\cite{chen2023wavmark} encode watermarks
through waveform-level perturbations learned and applied using a encoder-decoder pipeline, 
while approaches like AudioSeal~\cite{roman2024proactive} emphasize reliable and localised watermark detection through jointly trained generator-detector frameworks. 
Operating in the latent space of neural audio codecs generally enables high-fidelity watermarks deeply rooted
within the underlying core signal features~\cite{liu2023dear, wen2025watermark, li25g_interspeech}. 
Nonetheless, we have observed that the additive nature of existing schemes can leave watermarks undetectable in even modest levels of additive noise. 
We hence seek a different approach.


\section{Geometric principles}
\label{sec:principle}
Latent Secret Spin (LSS) is a blind speech watermarking framework which operates in the continuous latent space of a neural encoder.
LSS relies on a simple geometric idea whereby detectable changes in covariance are introduced using small orthogonal rotations in an anisotropic plane defined by principal components. 
In this section we describe the principle in its simplest form, independently of secret-key scheduling and distributed embedding strategies.

The underlying geometric principle is illustrated in Figure~\ref{fig:LSS_pipeline}.
Let $x$ denote an input speech waveform and let
\[
F = \mathcal{E}(x) \in \mathbb{R}^{n \times T}
\]
be the corresponding sequence of continuous latent features extracted using a neural encoder $\mathcal{E}$, where $n$ denotes the latent dimensionality and $T$ the number of analysis frames. 
To operate in a space with explicit and analytically tractable covariance structure, latent features are projected into a new space of principal components defined beforehand using PCA and a representative speech corpus. 
First, the same encoder $\mathcal{E}$ is applied to every utterance in the corpus.
We denote by $F_c$ the resulting corpus-level latent feature distribution.
Second, we define the global mean and covariance of $F_c$ as
\[
\mu = \mathbb{E}[F_c] \in \mathbb{R}^{n},
\qquad
\Sigma = \mathrm{Cov}(F_c) \approx U \Lambda U^\top,
\]
where the columns of $U$ are the principal components (i.e.\ the eigenvectors of $\Sigma$), where $U^\top U = I$, and where
\[
\Lambda = \mathrm{diag}(\lambda_1,\ldots,\lambda_n),
\qquad
\lambda_1 \geq \lambda_2 \geq \cdots > 0,
\]
contains the corresponding eigenvalues sorted in descending order, with $U$ ordered correspondingly. 
As illustrated in Figure~\ref{fig:LSS_pipeline}, latent features are then projected into the space defined by the principal components according to

\[
Z = U^\top(F-\mu).
\]

\begin{figure}[t]
    \centering
    \includegraphics[width=\columnwidth]{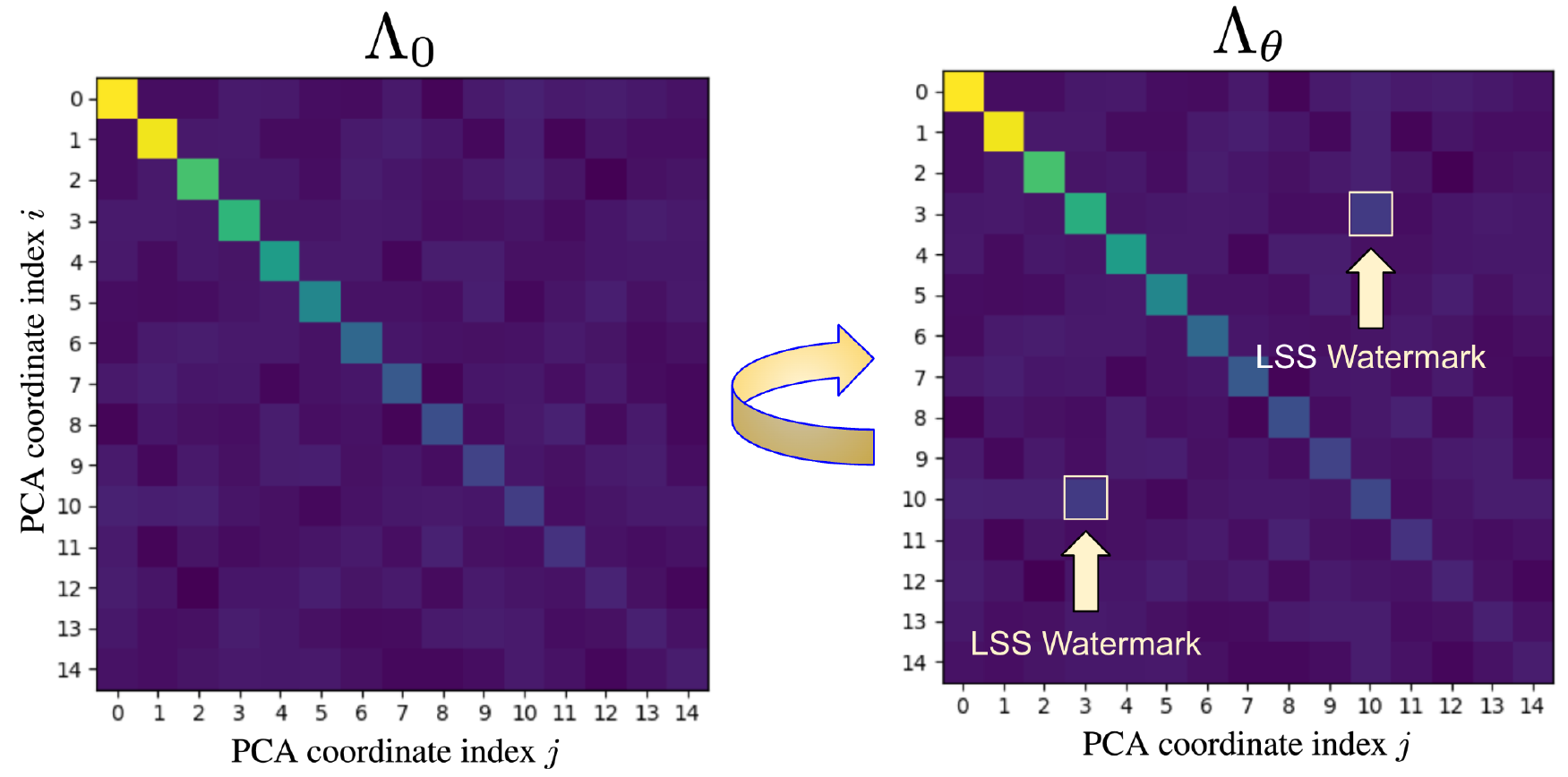}
    \caption{Illustration of the LSS watermarking geometric principle in PCA space.~$\Lambda_0$ denotes the covariance matrix before embedding, while~$\Lambda_\theta$ denotes the covariance matrix after a small rotation of angle $\theta = 0.18$ rad in the PCA plane $(i,j) = (3,10)$. The induced off-diagonal covariance terms correspond to the injected watermark. For readability, only a zoomed $15\times 15$ portion of the full PCA space is displayed.}
    \label{fig:LSS_toy}
    \vspace{-8pt}
\end{figure}

\begin{figure*}[t]
    \centering
    \includegraphics[width=\textwidth]
    {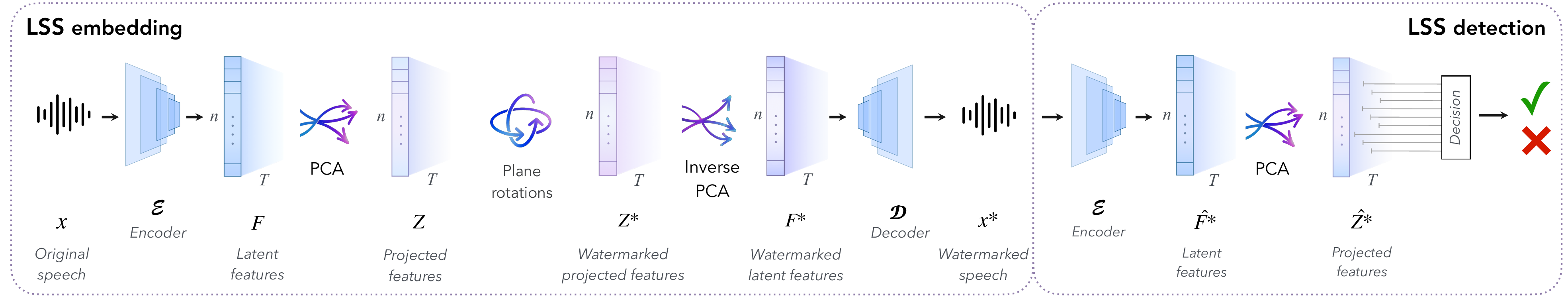}
    \caption{Overview of LSS embedding and detection pipelines. During embedding, the input signal $x$ is encoded into latent features $F$, projected to principal component space as $Z$, watermarked to obtain $Z^{\ast}$, mapped back to latent space $F^{\ast}$ and then decoded back to utterance $x^{\ast}$, watermarked. Trial signals are encoded and projected into the same space before watermark detection is performed.}
    \label{fig:LSS_pipeline}
    \vspace{-8pt}
\end{figure*}

The use of principal components 
provides a coordinate system in which the latent covariance is approximately diagonal, $\mathrm{Cov}(Z) \approx \Lambda,$

implying that latent dimensions are decorrelated. 
The variance of principal component axes is then, in general, distinct.\footnote{Unique eigenvalues are not strictly guaranteed. We have not observed repeated eigenvalues in practice.
} 
In this case the principal components space is anisotropic.

A watermark is then embedded by applying small orthogonal rotations to $Z$ in selected principal component planes. 
Consider a plane spanned by principal components $i,j$ such that $\lambda_i > \lambda_j$.
Before rotation, the covariance in plane $(i,j)$ is approximately
\[
\Sigma_0
=
\mathrm{Cov}([Z_i,Z_j])
\approx
\begin{bmatrix}
\lambda_i & 0\\
0 & \lambda_j
\end{bmatrix}.
\]
A rotation of angle $\theta>0$ defined by
\[
R(\theta)=
\begin{bmatrix}
\cos\theta & -\sin\theta\\
\sin\theta & \cos\theta
\end{bmatrix}
\]
is then applied. The covariance is then given by
\[
\Sigma_\theta = R(\theta)\,\Sigma_0\,R(\theta)^\top.
\]

\noindent The rotation introduces an off-diagonal covariance term

\begin{equation}
\begin{aligned}
\Delta \mathrm{Cov}_{ij}
&= (\lambda_i-\lambda_j)\sin\theta\cos\theta \\
&= \frac{1}{2}(\lambda_i-\lambda_j)\sin(2\theta),
\end{aligned}
\label{eq:footprint_exact}
\end{equation}
which, for small angles for which $\sin(2\theta)\approx 2\theta$, becomes
\begin{equation}
\Delta \mathrm{Cov}_{ij} \approx (\lambda_i-\lambda_j)\theta.
\label{eq:footprints}
\end{equation}
Thus, for $\lambda_i>\lambda_j$ and $|\theta| \ll 1$, $\Delta \mathrm{Cov}_{ij}>0$ for $\theta>0$, but $\Delta \mathrm{Cov}_{ij}<0$ for $\theta<0$.

This principle, illustrated in Fig.~1, is the fundamental mechanism behind LSS: small rotations in anisotropic planes induce an off-diagonal covariance term, the sign of which carries the sign (direction) of the rotation.

The rotation is applied in principal component space, yielding the watermarked representation $Z^{\ast}$.

Watermarked latent features are recovered by inverse projection:
\[
F^{\ast}=UZ^{\ast}+\mu.
\]
The decoder $\mathcal{D}$ is then used to reconstruct the watermarked speech waveform according to
\[
x^{\ast}=\mathcal{D}(F^{\ast}).
\]
The watermark can then be recovered trivially, even after $x^{\ast}$ is subject to various audio manipulations, perhaps applied by an adversary to render the watermark undetectable.
The watermarked speech signal $x^{\ast}$ can be re-encoded using the same encoder $\mathcal{E}$ and, with knowledge of the same principal components, an estimation $\hat{Z^{\ast}}$ can be recovered
and tested for the corresponding covariance signature.

In principle, the watermark introduced by the single, deterministic rotation in one principal component plane as described here is weak and insufficient.
An adversary could detect or recover such a trivial watermark relatively easily given a sufficient number of watermarked utterances.
As we shall now show, we build on this core principle by distributing rotations across multiple planes and temporal segments.
With knowledge of these planes and segments, local rotations can then be aggregated into statistically significant, detectable perturbations. 
In practice, rotations are applied in planes and segments selected according to a secret-key-controlled schedule, which guarantees undetectability without knowledge of the same secret-key.

\begin{figure}[t]
    \centering
    \includegraphics[width=\columnwidth]{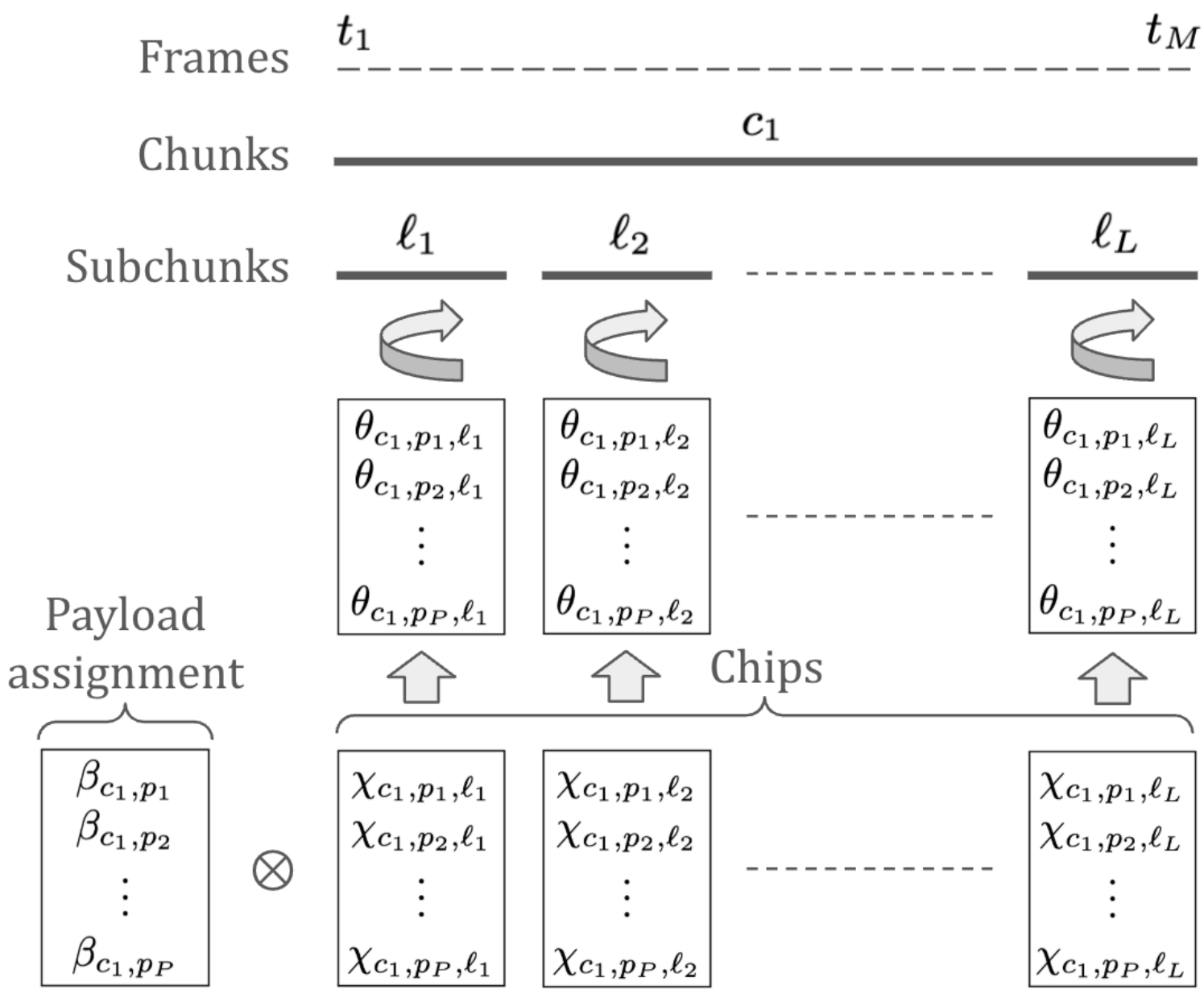}
    \caption{Watermark schedule within one chunk $c_1$. The chunk is partitioned into subchunks $\ell_1,\dots,\ell_L$. Each selected plane $p_1,\dots,p_P$ carries a payload sign $\beta_{c_1,p}$, while each subchunk is associated with a plane-dependent chip vector. Their product determines the local signed rotation angles $\theta_{c_1,p,\ell}$.}
    \label{fig:LSS_schedule}
    \vspace{-9pt}
\end{figure}

\begin{algorithm}[t]
\caption{LSS Watermark Embedding}
\label{alg:embedding}
\begin{algorithmic}
\State \textbf{Input:} Input signal $x$, PCA basis $(U,\mu)$, encoder $\mathcal{E}$, decoder $\mathcal{D}$, key $K$, nonce $N$, rotation magnitude $\theta$, payload bits $\mathbf{b}=(b_1,\dots,b_B)$
\State \textbf{Output:} Watermarked signal $x^{\ast}$
\Statex 
\State $F \leftarrow \mathcal{E}(x)$
\State $Z \leftarrow U^\top(F-\mu)$
\State Convert payload bits $\mathbf{b}$ into signed circular list $\boldsymbol{\beta}=(\beta_1,\dots,\beta_{B})$
\State Partition $Z$ into $C$ chunks $[c_1,\dots,c_{C}]$
\Statex \textcolor{gray}{\textit{// Each chunk $c$ is associated to a list of planes generated pseudo-randomly}}
\For{each chunk $c$ in $[c_1,\dots,c_{C}]$}
    \State $[p_1, \dots,p_{P}]_c \leftarrow \text{pseudo\_random}(K,N)$
    
    \Statex \textcolor{gray}{\textit{// Each plane $p$ encodes a bit from $\boldsymbol{\beta}$}}
    \For{each plane $p$ in $[p_1, \dots,p_{P}]_c$}
        \State $\beta_{c,p} \leftarrow \text{get\_next\_bit}(\boldsymbol{\beta})$
        \State Divide $c$ into $L$ subchunks $[\ell_1,\dots,\ell_{L}]$
        \Statex \textcolor{gray}{\textit{// One bit is scattered across subchunks.}}
        \Statex \textcolor{gray}{\textit{// Each subchunk is associated to a chip $\chi$ generated pseudo-randomly}}
        \For{each subchunk $\ell$ in $[\ell_1,\dots,\ell_{L}]$}
            \State $\chi_{c,p,\ell} \leftarrow \text{pseudo\_random}(K,N)$
            \State $\theta^{\ast}_{c,p,l} \leftarrow \beta_{c,p}  \cdot \chi_{c,p,\ell}  \cdot \theta$
            \State $Z^{\ast} \leftarrow$ Rotate plane $p$ by $\theta^{\ast}_{c,p,l}$ over each $t$ in subchunk $\ell$ 
        \EndFor
    \EndFor
\EndFor
\State $F^{\ast} \leftarrow UZ^{\ast} + \mu$
\State $x^{\ast} \leftarrow \mathcal{D}(F^{\ast})$
\State \Return $x^{\ast}$
\end{algorithmic}
\end{algorithm}
\vspace{-5pt}

\section{Latent Secret Spin}
\label{sec:implementation}

We now describe the practical implementation of LSS, including watermark embedding, watermark detection, and the keyed pseudo-random schedule.

\subsection{Watermark embedding}
\label{subsec:embedding}

The embedding procedure is applied to $Z$ across temporal chunks and subchunks, and across a set of geometric planes. 
The process is described in the following, illustrated in Fig.~\ref{fig:LSS_schedule} and summarised in Algorithm~\ref{alg:embedding}.

First, the sequence of projected latent features $Z$ is partitioned into $C$ chunks
$c_1,\dots,c_C$ each of length $M$ frames $t_1,\dots,t_M$ where $t$ denotes the frame index. For each chunk $c$,  we select according to the pseudo-random schedule a set of principal component planes $p_1,\dots,p_P$ where each plane $p$ corresponds to a pair of principal directions $(i_p,j_p)$. Each chunk is further subdivided into $L$ subchunks $\ell_1,\dots,\ell_L$.

The watermark is encoded as a $B$-bit sequence referred to as the \emph{payload}.
In practice, it is represented as a vector $\boldsymbol{\beta}\in\left\{\pm 1\right\}^B$, where bit values $1$ become $+1$ and bit values $0$ become $-1$.
Signed bits from $\boldsymbol{\beta}$ are embedded into $Z$ sequentially; namely,
each signed payload bit $\beta_{c,p}\in\{\pm1\}$ is embedded through rotations in a plane $p$ across the chunk $c$.
When all bits have been assigned, $\boldsymbol{\beta}$ is repeated cyclically.

The embedding of a bit in a plane is distributed over temporal subchunks:

for each selected plane $p$ and each subchunk $\ell$, we determine from the schedule a pseudo-random chip

\[
\chi_{c,p,\ell}\in\{\pm1\},
\]
which controls the direction of the local rotation in that plane and subchunk (see Fig.~\ref{fig:LSS_schedule}).

The resulting signed rotation for chunk $c$, plane $p$ and sub-chunk $l$ is hence given by
\[
\theta^{\ast}_{c,p,\ell}
=
\beta_{c,p}\cdot \chi_{c,p,\ell}\cdot \theta 
\]

where $\theta > 0$ is a fixed rotation angle chosen as hyperparameter.
The value of $\theta^{\ast}_{c,p,\ell}$ is either $+\theta$ or $-\theta$, where the sign is determined by the signs of the bit and the chip.
Hence, within each subchunk $\ell$,
chip $\chi_{c,p,\ell}$ acts to modulate the 
signed payload carrier $\beta_{c,p}$. 
For each chunk $c$ and each plane $p=(i_p, j_p)$,
each feature frame $Z(t)$ has its $i_p$ and $j_p$ components rotated according to

\[
\begin{bmatrix}
Z^{\ast}_{i_p}(t)\\
Z^{\ast}_{j_p}(t)
\end{bmatrix}
=
R\left(\theta^{\ast}_{c,p,\ell}\right)
\begin{bmatrix}
Z_{i_p}(t)\\
Z_{j_p}(t)
\end{bmatrix},
\qquad t\in \ell.
\]
Thus, the watermark is distributed across multiple geometric planes and temporal segments. 
After the set of rotations has been applied, the modified representation $Z^{\ast}$ is projected back to the original latent space $F^{\ast}$ and then decoded, yielding the watermarked waveform $x^{\ast}$.

\subsection{Watermark detection}
\label{subsec:detection}
A summary of the detection procedure is presented in Algorithm~\ref{alg:detection}.
Using the same neural encoder $\mathcal{E}$, the detector first projects a trial waveform $x'$ into latent space $F'$.  Then, according to the same principal components
used for initial encoding, an approximation of potentially watermarked features $Z'$ is recovered.
The representation is then processed according to the same watermark schedule, with chunks $c_1,\dots,c_C$, subchunks $\ell_1,\dots,\ell_L$, planes $p_1,\dots,p_P$ and chips $\chi_{c,p,\ell}$. 
Then, for each chunk $c$, plane $p$, and each subchunk $\ell$, the detector computes the  
normalised covariance 
\[
C_{c,p,\ell}
=
\frac{1}{|\ell|{\sqrt{\lambda_{i_p}\lambda_{j_p}}}}
\sum_{t\in \ell}
\bigl(Z'_{i_p}(t)-\bar Z'_{i_p,\ell}\bigr)
\bigl(Z'_{j_p}(t)-\bar Z'_{j_p,\ell}\bigr),
\]
where $p$ is the plane spanning principal components $(i_p,j_p)$, $\bar Z'_{i_p,\ell}$ and $\bar Z'_{j_p,\ell}$ denote the local mean for each component over sub-chunk $\ell$, and $|\ell|$ is the number of frames per sub-chunk. 
Bits $\beta_{c,p}$ and chips $\chi_{c,p,\ell}$ are derived from the schedule in the same way as in the embedding process. They are accumulated, along with each corresponding $C_{c,p,\ell}$,
into a detection score
\begin{equation}
S
=
\sum_c \sum_p \sum_{\ell}
\beta_{c,p}\,\chi_{c,p,\ell}\,C_{c,p,\ell}\
\label{eq:global_score}
\end{equation}
where the bit and the chip compensate for the sign modulation introduced during embedding. 
In this case, signed covariance terms sum constructively and the score $S$ is
positive.
In the case of a different schedule produced by an incorrect key, a mismatched payload, or an absent watermark, covariance terms
are equally likely to have positive or negative sign:
they will sum destructively in~\eqref{eq:global_score}, producing a lower score.
Binary detection decisions are then produced by a comparison between $S$ and a decision threshold.

\subsection{Keyed pseudo-random schedule generation}

LSS watermark schedules are generated pseudo-randomly using a secret key $K$ and a utterance-specific nonce $N$ and are shared by the encoder and decoder.  
This keyed schedule
determines the selection of planes $p$ and the chips $\chi_{c,p,\ell}$ during embedding. 

The detector evaluates the covariance terms in the same sequence of planes and chip signs as during embedding, reproduced deterministically from the pseudo-random function using the same $K,N$ pair.
Without it, the plane and chip sequence is different: therefore, the sign of $C_{c,p,\ell}$ oscillates randomly between positive and negative,
and the contributions in \eqref{eq:global_score} cancel out on average. This results in $S\approx0$, preventing watermark detection by any authorized party who does not possess both $K$ and $N$.


\begin{algorithm}[t]
\caption{LSS Watermark Detection}
\label{alg:detection}
\begin{algorithmic}
\State \textbf{Input:} Investigated signal $x'$, PCA basis $(U,\mu)$, encoder $\mathcal{E}$, key $K$, nonce $N$, payload $\boldsymbol{\beta}$, decision threshold $\tau$
\State \textbf{Output:} Decision if $\boldsymbol{\beta}$ is watermarked into $x'$ 
\Statex 
\State $S \leftarrow 0$ \ \ \textcolor{gray}{\textit{// Watermark score tracker}}
\State $F' \leftarrow \mathcal{E}(x')$
\State $Z' \leftarrow U^\top(F'-\mu)$
\Statex \textcolor{gray}{\textit{// Loop as in the embedding process}}
\State Partition $Z'$ into $C$ chunks $[c_1,\dots,c_{C}]$
\For{each chunk $c$ in $[c_1,\dots,c_{C}]$}
    \State $[p_1, \dots,p_{P}]_c \leftarrow \text{pseudo\_random}(K,N)$
    \For{each plane $p$ in $[p_1, \dots,p_{P}]_c$}
        \State $\beta_{c,p} \leftarrow \text{get\_next\_bit}(\boldsymbol{\beta})$
        \State Divide $c$ into $L$ subchunks $[\ell_1,\dots,\ell_{L}]$
        \For{each subchunk $\ell$ in $[\ell_1,\dots,\ell_{L}]$}
            \State $\chi_{c,p,\ell} \leftarrow \text{pseudo\_random}(K,N)$
            \State \textcolor{gray}{\textit{// Compute covariance term of plane $p$ over individual frames of subchunk $\ell$}}
            \State $i_p, j_p \leftarrow \text{pair of principal components defining } p$ 
            \[
            C_{c, p ,\ell} \leftarrow \frac{\sum_{t\in\ell}
            \bigl(Z'_{i_p}(t)-\bar Z'_{i_p,\ell}\bigr)\bigl(Z'_{j_p}(t)-\bar Z'_{j_p,\ell}\bigr)}{\lvert \ell \rvert \sqrt{\lambda_{i_p}\lambda_{j_p}}} 
            \]
            \State $S \leftarrow S + \chi_{c, p, \ell} \cdot \beta_{c,p} \cdot C_{c,p,\ell}$
        \EndFor
    \EndFor
\EndFor
\State \Return \textit{True} if $S>\tau$, \textit{False} otherwise
\end{algorithmic}
\end{algorithm}

\section{Experimental Setup}
\label{sec:setup}

We evaluate the effectiveness of the proposed LSS framework in both in-domain and out-of-domain settings. To assess robustness, we apply various audio manipulations to watermarked speech before detection. 

\vspace{-2pt}
\subsection{Datasets}
\label{subsec:datasets}
We evaluate the proposed LSS framework using two different speech datasets: VoxPopuli~\cite{wang-etal-2021-voxpopuli} and ASVspoof~5~\cite{WANG2026101825}. For both, we create two speaker-disjoint partitions.  
The first is used to derive the set of principal components, while the second is used for evaluation.
We use the English subset of the $100$k-hour unlabelled partition of VoxPopuli.
Principal components are estimated using approximately $100$k samples, while evaluation is performed using a second set of $10$k utterances for embedding and detection experiments.
We use data sourced from the ASVspoof\,5 evaluation partition from which we select uncompressed bona fide utterances. 
From $\approx35$k utterances, we set aside $10$k samples for evaluation and use the remaining $25$k utterances to estimate principle components. 

Following~\cite{roman2024proactive}, all utterances are standardised to a duration of $10$~s. Shorter recordings are circularly padded to avoid introducing zeros that would otherwise distort latent covariance statistics, while longer utterances are truncated.


\subsection{Configuration}
\label{subsec: codec}

Encoder $\mathcal{E}$ and decoder $\mathcal{D}$ are pre-trained $24\,\text{kHz}$ implementations of EnCodec~\cite{defossez2023high}\footnote{We resample 16 kHz data to 24 kHz.} operating at a target bandwidth of $6.0\,\text{kbps}$.
Feature representations have dimensionality $n=128$ and are produced at a frame rate of $75\,\text{Hz}$.

We use chunk sizes of $M=32$ frames ($\approx427\,\mathrm{ms}$), subchunk sizes of $L=8$ frames ($\approx107\,\mathrm{ms}$), and $P=24$ planes.

We achieve reliable detection and imperceptible watermarking using 
rotations of $\theta=0.18\,\mathrm{rad}$. 

Source code\footnote{Code and sample utterances are available at \url{https://github.com/eurecom-asp/lss}} is available and can be used to reproduce all LSS results reported in this paper.  
Those of other techniques are reported from~\cite{roman2024proactive}.

\vspace{-2pt}
\subsection{Audio manipulations}
\label{subsec:manipulations}

We apply seven manipulations to watermarked utterances $x^{\ast}$ before detection, namely sixth-order Butterworth lowpass, highpass, and bandpass filters, MP3 compression at bitrates from $12$ to $256\,\text{kbps}$, resampling from $24\,\text{kHz}$ to $16\,\text{kHz}$ and back, and additive white Gaussian and $1/f$ pink noise at SNRs from $5$ to $20\,\text{dB}$.

\vspace{-2pt}
\subsection{Evaluation metrics} 
\label{subsec: metrics}
Following~\cite{roman2024proactive}, we report AUC-ROC of watermark detection. We do not compute accuracy as it does not reflect the overall strength of the method, but rather its performance at a specific, use case-dependent decision threshold.

In addition, we estimate imperceptibility using the two-sided objective perceptual evaluation of speech quality metric with a wideband setting (PESQ-WB).

\section{Results}
\label{sec:results}

In the following we present an analysis of detection performance, robustness and estimated perceptual quality.

\vspace{-2pt}
\subsection{Detection Performance}
\label{subsec: detection_results}

\begin{figure*}[t]
\centering

\begin{subfigure}{0.245\textwidth}
    \centering
    \includegraphics[width=\linewidth]{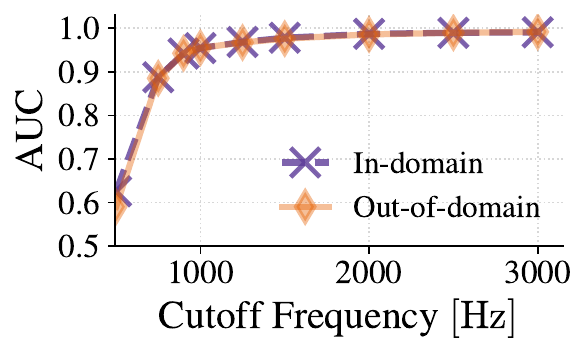}
    \caption{Low-pass filter}
\end{subfigure}
\hfill
\begin{subfigure}{0.245\textwidth}
    \centering
    \includegraphics[width=\linewidth]{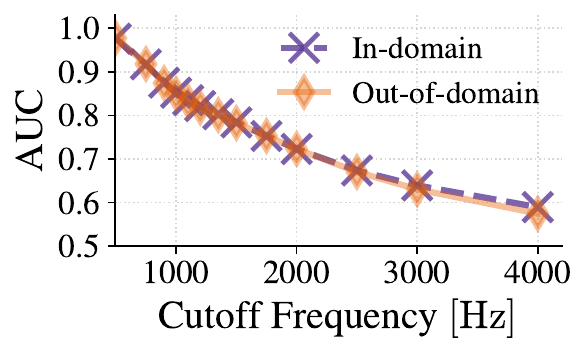}
    \caption{High-pass filter}
\end{subfigure}
\hfill
\begin{subfigure}{0.245\textwidth}
    \centering
    \includegraphics[width=\linewidth]{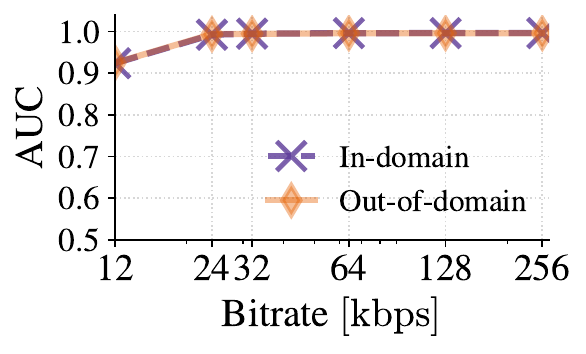}
    \caption{MP3 Compression}
\end{subfigure}
\hfill
\begin{subfigure}{0.245\textwidth}
    \centering
    \includegraphics[width=\linewidth]{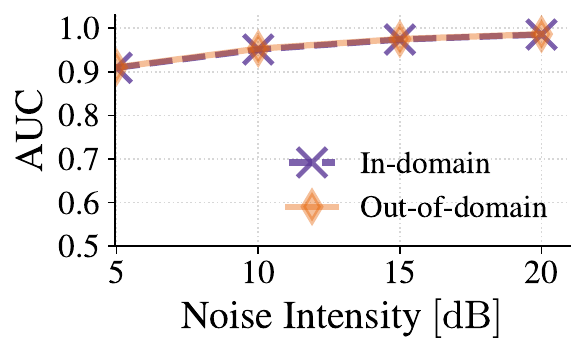}
    \caption{Additive White Noise}
\end{subfigure}

    \caption{Detection accuracy as a function of manipulation intensity for four types of signal distortions.  Purple curves denote the in-domain scenario (T2) and orange curves the out-of-domain setting (T3).}
\label{fig:robustness_plots}
\end{figure*}

\begin{figure}[!t]
\centering

\begin{subfigure}{0.48\columnwidth}
    \centering
    \includegraphics[width=\linewidth]{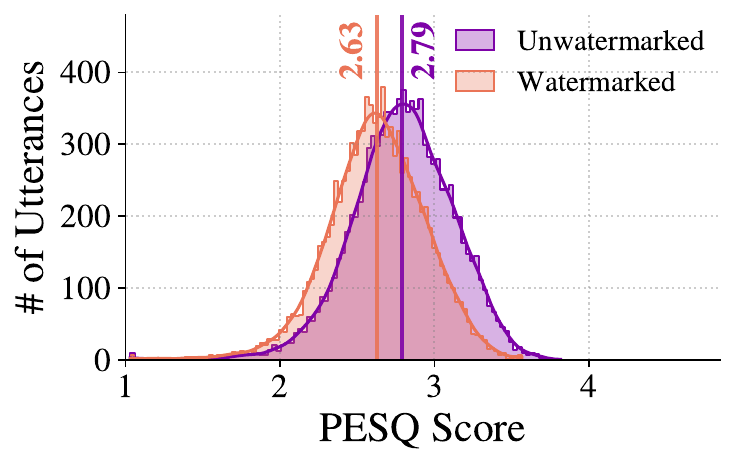}
    \caption{In-domain scenario T2.}
\end{subfigure}
\hfill
\begin{subfigure}{0.48\columnwidth}
    \centering
    \includegraphics[width=\linewidth]{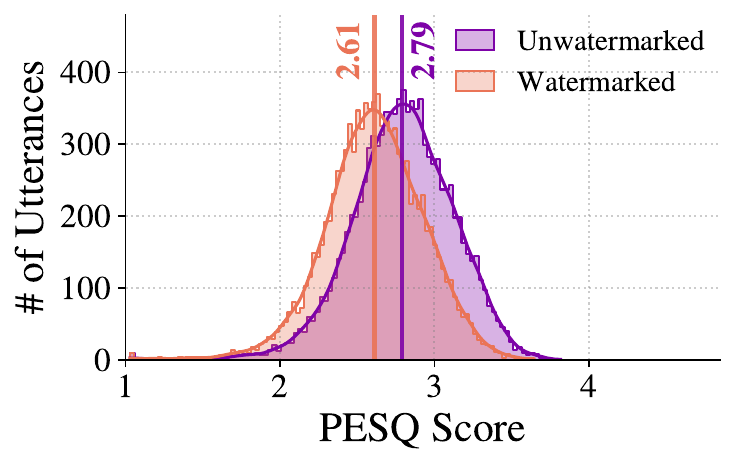}
    \caption{Out-of-domain scenario T3.}
\end{subfigure}

\caption{Distribution of PESQ scores for codec-reconstructed speech before (purple) and after (orange) watermark embedding. The annotated values indicate the average PESQ computed over the full evaluation set.}
\label{fig:pesq_scores}
\vspace{-10pt}
\end{figure}

Detection results are reported in Table~\ref{tab:clean_detection} for both in-domain and out-of-domain conditions, depending on whether principal components are estimated and evaluated on the same dataset or on different datasets. In all cases, detection performance remains strong when the correct key is used, with consistently high AUC values. The close agreement between in-domain and out-of-domain results indicates that the principal component representation generalises well across datasets and that detection does not rely on dataset-specific characteristics. By contrast, when an incorrect key is used, performance collapses to random guess level (AUC $\approx 0.5$), confirming that reliable detection depends on knowledge of the correct secret key.


LSS achieves strong detection performance across most distortions, as shown in Table~\ref{tab:LSS}. In particular, under common channel effects such as MP3 compression, resampling, additive noise at $20\,\mathrm{dB}$, and bandwidth filtering, performance remains close to the clean condition, indicating good robustness to realistic transmission degradations. For contextual reference, AudioSeal~\cite{roman2024proactive} achieves a slightly higher average AUC overall (97.2\% versus 95.6\% for LSS), while LSS remains close on average and outperforms AudioSeal under lowpass filtering at $f_c=1\,\mathrm{kHz}$. These manipulations were chosen to cover a range of common post-processing operations, including spectral filtering, compression, resampling, and additive noise.

To further analyse robustness, Fig.~\ref{fig:robustness_plots} reports detection accuracy as a function of manipulation intensity for several distortions. Results are shown for both matched (T2) and mismatched (T3) scenarios defined in Table~\ref{tab:clean_detection}. Across all conditions, the two curves remain closely aligned, indicating that detection does not rely on dataset-specific cues but instead exploits stable statistical properties of the codec latent representation. This behaviour further suggests that LSS generalises well under distribution shifts combined with signal degradation.

\begin{table}[t]
\centering
\setlength{\tabcolsep}{4pt}
\renewcommand{\arraystretch}{0.99}
\footnotesize
\caption{\textbf{Detection performance on clean watermarked speech} for all dataset scenarios. Results are shown for the correct key only; incorrect-key detection is at random guess level (AUC $\approx 50.0\%$).}
\resizebox{\columnwidth}{!}{
\begin{tabular}{@{}l l l l c@{}}
\toprule
\textbf{Scen.} & \textbf{Dom.} & \textbf{Training} & \textbf{Evaluation} & \textbf{AUC (\%)} \\
\midrule
T1 & In  & VoxPopuli-100k & VoxPopuli-100k & 99.6 \\
T2 & In  & ASVspoof 5      & ASVspoof5      & 99.7 \\
T3 & Out & VoxPopuli-100k & ASVspoof5      & 99.7 \\
T4 & Out & ASVspoof 5      & VoxPopuli-100k & 99.3 \\
\bottomrule
\end{tabular}}
\label{tab:clean_detection}
\end{table}

\begin{table}[!t]
    \centering
    \footnotesize
    \setlength{\tabcolsep}{3pt}
    \caption{\textbf{Detection performance under non-malicious audio manipulations.} Results are reported for LSS and AudioSeal on configuration T1 of Table \ref{tab:clean_detection}.}
    \label{tab:LSS}
    \begin{tabularx}{\columnwidth}{@{} l l 
    >{\centering\arraybackslash}X 
    >{\centering\arraybackslash}X @{}}
        \toprule
        & & \textbf{LSS} & \textbf{AudioSeal} \\
        \textbf{Manipulation} & \textbf{Configuration} 
        & \textbf{AUC (\%)} & \textbf{AUC (\%)} \\
        \midrule
        None         & -- & 99.6 & 100.0 \\
        Lowpass      & $f_c=1\,\mathrm{kHz}$ 
                     & 96.5 & \textbf{67.8} \\
        Lowpass      & $f_c=1.5\,\mathrm{kHz}$  
                     & 98.1 & 100.0 \\
        Highpass     & $f_c=1\,\mathrm{kHz}$  
                     & 87.3 & 100.0 \\
        Highpass     & $f_c=1.5\,\mathrm{kHz}$  
                     & 80.3 & 100.0 \\
        Bandpass     & $500\,\mathrm{Hz}$\,--\,$5\,\mathrm{kHz}$ 
                     & 97.4 & 100.0 \\
        MP3          & $32\,\mathrm{kbps}$ 
                     & 99.5 & 100.0 \\
        Resample     & $[24\rightarrow16\rightarrow24]\,\mathrm{kHz}$ 
                     & 99.7 & 100.0 \\
        White Noise  & SNR = $5\,\mathrm{dB}$ 
                     & 94.8 & 99.8 \\
        White Noise  & SNR = $20\,\mathrm{dB}$ 
                     & 99.3 & 100.0 \\
        Pink Noise   & SNR = $5\,\mathrm{dB}$ 
                     & 95.6 & 99.9 \\
        Pink Noise   & SNR = $20\,\mathrm{dB}$ 
                     & 99.4 & 100.0 \\
        \midrule
        Average      &  & 95.6 & 97.2 \\
        \bottomrule
    \end{tabularx}
    \vspace{-8pt}
\end{table}

\vspace{-2pt}
\subsection{Estimated perceptual quality}
\label{subsec: quality_eval}
Figure~\ref{fig:pesq_scores} shows PESQ distributions for the in-domain scenario T2 and the out-of-domain scenario T3 defined in Table~\ref{tab:clean_detection}. In both cases, distributions for unwatermarked and watermarked utterances largely overlap, indicating only modest quality degradation due to LSS watermarking. Across all configurations, the average drop is below $\Delta\mathrm{PESQ}=0.2$.

\section{Discussion}
\label{sec:discussion}
LSS shows that blind speech watermarking can be achieved effectively in codec latent space by exploiting the anisotropic structure of PCA representations. The method provides reliable keyed detection, generalises well across datasets, and remains robust under common signal manipulations while preserving perceptual quality. 

Key strengths of LSS are its simplicity, interpretability, and high degree of flexibility. Unlike fully learned watermarking systems, it does not rely on a dedicated trained embedder-detector pair, but instead operated through small orthogonal rotations in selected PCA planes. The resulting covariance changes are predictable and analytically tractable, making the method easier to reason about in terms of robustness, security and failure modes.

This design also allows the behaviour of LSS to be adjusted through tuning of parameters such as number of planes $P$ or rotation magnitude $\theta$, without retraining. More broadly, this statistical formulation provides a complementary perspective to fully learned watermarking approaches and while not being tied to a specific encoder, suggesting that it could in principle be applied to other suitable latent representation.

Results indicate that LSS remains stable across heterogeneous manipulations, including the residual vector quantization naturally introduced by EnCodec. 

In contrast to approaches such as Audioseal~\cite{roman2024proactive}, where robustness is learned through exposure to distortions during training, LSS exhibits structural robustness: since watermarking is performed directly in the codec latent space, the watermark is inherently subjected to the same quantisation and compression operations as the underlying representation. Robustness to neural codec compression therefore arises naturally from the embedding space rather than being learned as an external invariance.

Although high-pass filtering remains the most challenging condition, LSS degrades more gracefully than competing methods~\cite{roman2024proactive}, whose performance varies more strongly across filtering manipulations. This more predictable behaviour is attractive in practical scenarios, where signal post-processing is typically unknown. 

This is particularly encouraging because LSS was not extensively optimised: 
the reported configuration was chosen primarily to validate the geometric principle, suggesting that further improvements may be achieved through tuning of the embedding schedule and rotation parameters.

LSS also occupies an interesting position with respect to payload design:
the embedding depends on payload-specific sign assignments, so that the hidden message structure directly shapes the watermark pattern.

LSS does not impose a fixed payload length a priori. In contrast, in AudioSeal the payload capacity is defined during training and cannot be increased afterward without retraining the model. In addition, unlike AudioSeal, LSS restricts watermark detection to authorised parties holding the secret key: even if the decoder weights are publicly disclosed, detection remains infeasible without access to the key.
While in this work we focus on watermark detection, future investigations will explore the possibility of explicit payload recovery.

The presented study has some limitations. Experiments are restricted to bona fide speech and a fixed codec configuration, while robustness is evaluated under common, non-malicious manipulations rather than stronger, adaptive attacks. 
Since watermarks  are distributed at the chunk level, splicing and other, similar temporal manipulations  will impact upon detection reliability and require further consideration.
In addition, perceptual quality is estimated through only objective metrics rather than subjective listening tests. 
We leave these limitations and the broader design space of LSS to be examined more thoroughly in future work, particularly under stronger attack models and across alternative latent representations.

\section{Conclusions}
\label{sec:conclusion}
We introduced LSS, a blind speech watermarking method based on small orthogonal rotations in structured latent spaces. By exploiting the anisotropic geometry of PCA representations, LSS induces a detectable covariance signature that enables reliable keyed detection without requiring a trained embedder or detector.
Experiments show that LSS is robust, perceptually transparent, and able to generalise across datasets. Results show that geometric operations in codec latent spaces can provide a stronger and interpretable alternative to fully learned watermarking approaches.
LSS therefore offers a promising direction for robust and imperceptible speech watermarking, while paving the way for future work using stronger attack models, richer payload embedding and recovery, and improved schedule design.

\section{Acknowledgements}
\label{sec:acknow}
This work was supported by the COMPROMIS project (ANR22-PECY-0011) funded by a French government grant managed by the Agence Nationale de la Recherche under the France 2030 program.
\newpage
\bibliographystyle{IEEEtran}
\bibliography{Odyssey2026_BibEntries}

@inproceedings{
cao2025watermarking,
title={{Watermarking for {AI} Content Detection: A Review on Text, Visual, and Audio Modalities}},
author={Lele Cao},
booktitle={The 1st Workshop on GenAI Watermarking},
year={2025},
}

@article{charfeddine2022audio,
  title={{Audio watermarking for security and non-security applications}},
  author={Charfeddine, Maha and Mezghani, Eya and Masmoudi, Salma and Amar, Chokri Ben and Alhumyani, Hesham},
  journal={IEEE Access},
  volume={10},
  pages={12654--12677},
  year={2022},
  publisher={IEEE}
}

@article{lie2006robust,
  title={Robust and high-quality time-domain audio watermarking based on low-frequency amplitude modification},
  author={Lie, Wen-Nung and Chang, Li-Chun},
  journal={IEEE transactions on multimedia},
  volume={8},
  number={1},
  pages={46--59},
  year={2006},
  publisher={IEEE}
}

@article{hua2016twenty,
  title={Twenty years of digital audio watermarking—a comprehensive review},
  author={Hua, Guang and Huang, Jiwu and Shi, Yun Q and Goh, Jonathan and Thing, Vrizlynn LL},
  journal={Signal processing},
  volume={128},
  pages={222--242},
  year={2016},
  publisher={Elsevier}
}

@article{faundez2010speech,
  title={Speech watermarking: an approach for the forensic analysis of digital telephonic recordings},
  author={Faundez-Zanuy, Marcos and Lucena-Molina, Jose J and Hagm{\"u}ller, Martin},
  journal={Journal of forensic sciences},
  volume={55},
  number={4},
  pages={1080--1087},
  year={2010},
  publisher={Wiley Online Library}
}

@book{barni2004watermarking,
  title={Watermarking systems engineering: enabling digital assets security and other applications},
  author={Barni, Mauro and Bartolini, Franco},
  year={2004},
  publisher={Crc Press}
}

@inproceedings{liu2023dear,
author = {Liu, Chang and Zhang, Jie and Fang, Han and Ma, Zehua and Zhang, Weiming and Yu, Nenghai},
title = {{DeAR: a deep-learning-based audio re-recording resilient watermarking}},
year = {2023},
isbn = {978-1-57735-880-0},
publisher = {AAAI Press},
url = {https://doi.org/10.1609/aaai.v37i11.26550},
doi = {10.1609/aaai.v37i11.26550},
booktitle = {Proceedings of the Thirty-Seventh AAAI Conference on Artificial Intelligence and Thirty-Fifth Conference on Innovative Applications of Artificial Intelligence and Thirteenth Symposium on Educational Advances in Artificial Intelligence},
articleno = {1481},
numpages = {9},
series = {AAAI'23/IAAI'23/EAAI'23}
}

@article{wen2025watermark,
  author={Wen, Shuangbing and Zhang, Qishan and Hu, Tao and Li, Jun},
  journal={IEEE Signal Processing Letters}, 
  title={{Robust Audio Watermarking Against Manipulation Attacks Based on Deep Learning}}, 
  year={2025},
  volume={32},
  number={},
  pages={126-130},
  doi={10.1109/LSP.2024.3501285}}

@inproceedings{li25g_interspeech,
  title     = {{VoiceMark: Zero-Shot Voice Cloning-Resistant Watermarking Approach Leveraging Speaker-Specific Latents}},
  author    = {Haiyun Li and Zhiyong Wu and Xiaofeng Xie and Jingran Xie and Yaoxun Xu and Hanyang Peng},
  year      = {2025},
  booktitle = {{Interspeech 2025}},
  pages     = {5108--5112},
  doi       = {10.21437/Interspeech.2025-575},
  issn      = {2958-1796},
}

@article{murata2011audio,
  title={{Audio watermarking using Wavelet Transform and Genetic Algorithm for realizing high tolerance to MP3 compression}},
  author={Murata, Shinichi and Yoshitomi, Yasunari and Ishii, Hiroaki},
  journal={Journal of Information Security},
  volume={2},
  number={3},
  pages={99},
  year={2011},
  publisher={Scientific Research Publishing}
}

@article{dhar2013dwt,
  title={{A DWT-DCT-based audio watermarking method using singular value decomposition and quantization}},
  author={Dhar, Pranab Kumar and Shimamura, Tetsuya},
  journal={Journal of Signal Processing},
  volume={17},
  number={3},
  pages={69--79},
  year={2013},
  publisher={Research Institute of Signal Processing, Japan}
}

@inproceedings{celik2005pitch,
  author={Celik, M. and Sharma, G. and Tekalp, A.M.},
  booktitle={Proceedings. (ICASSP '05). IEEE International Conference on Acoustics, Speech, and Signal Processing, 2005.}, 
  title={Pitch and duration modification for speech watermarking}, 
  year={2005},
  volume={2},
  number={},
  pages={ii/17-ii/20 Vol. 2},
  doi={10.1109/ICASSP.2005.1415330}}

@inproceedings{chawla2018ARO,
  title={{A Review of DWT and PCA based Digital Watermarking Schemes}},
  author={Nidhi Chawla and Vikram Singh},
  year={2018},
  url={https://api.semanticscholar.org/CorpusID:212595206},
}

@inproceedings{Tonge2014ASO,
  title={{A Survey of Digital Watermarking Techniques}},
  author={Madhuri Tonge and Anshu Gupta and Rajiv Gandhi and Proudyogiki Vishwavidyalya},
  year={2014},
  url={https://api.semanticscholar.org/CorpusID:115557554}
}

@inproceedings{wang2018speech,
  author={Wang, Shengbei and Yuan, Weitao and Wang, Jianming and Unoki, Masashi},
  booktitle={2018 IEEE International Conference on Acoustics, Speech and Signal Processing (ICASSP)}, 
  title={{Speech Watermarking Based on Robust Principal Component Analysis and Formant Manipulations}}, 
  year={2018},
  volume={},
  number={},
  pages={2082-2086},
  doi={10.1109/ICASSP.2018.8462356}}

@article{wang2020secure,
author = {Wang, Shengbei and Wang, Chao and Yuan, Weitao and Wang, Lin and Wang, Jianming},
title = {{Secure echo-hiding audio watermarking method based on improved PN sequence and robust principal component analysis}},
journal = {IET Signal Processing},
volume = {14},
number = {4},
pages = {229-242},
doi = {https://doi.org/10.1049/iet-spr.2019.0376},
url = {https://ietresearch.onlinelibrary.wiley.com/doi/abs/10.1049/iet-spr.2019.0376},
eprint = {https://ietresearch.onlinelibrary.wiley.com/doi/pdf/10.1049/iet-spr.2019.0376},
year = {2020}
}

@article{kahdim2023principal,
  title={Principal component analysis for zero watermarking technique},
  author={Kahdim, Saja Abdulameer and Abduldaim, Areej M},
  journal={Comput Sci},
  volume={18},
  number={1},
  pages={85--97},
  year={2023}
}

@article{yang2025zero,
  title={{Zero Watermarking Algorithm for Hyperspectral Remote Sensing Images Considering Spectral and Spatial Features}},
  author={Yang, Bingbing and Yan, Haowen and Zhang, Liming and Yan, Qingbo and Hou, Zhaoyang and Wang, Xiaolong and Xu, Xinyu},
  journal={IEEE Journal of Selected Topics in Applied Earth Observations and Remote Sensing},
  year={2025},
  publisher={IEEE}
}

@article{saboori2014ANM,
  title={{A new method for digital watermarking based on combination of DCT and PCA}},
  author={Arash Saboori and Seyed Amir Hossein Hosseini},
  journal={2014 22nd Telecommunications Forum Telfor (TELFOR)},
  year={2014},
  pages={521-524},
  url={https://api.semanticscholar.org/CorpusID:9023750}
}

@article{sinha2011digital,
  title={{Digital video watermarking using discrete wavelet transform and principal component analysis}},
  author={Sinha, Sanjana and Bardhan, Prajnat and Pramanick, Swarnali and Jagatramka, Ankul and Kole, Dipak K and Chakraborty, Aruna},
  journal={International Journal of Wisdom Based Computing},
  volume={1},
  number={2},
  pages={7--12},
  year={2011}
}

@inproceedings{hien2003pca,
  title={{PCA based digital watermarking}},
  author={Hien, Thai D and Chen, Yen-Wei and Nakao, Zensho},
  booktitle={International Conference on Knowledge-Based and Intelligent Information and Engineering Systems},
  pages={1427--1434},
  year={2003},
  organization={Springer}
}

@article{wang2000watermarking,
  title={{Watermarking based on principal component analysis}},
  author={Wang, Shuo-zhong},
  journal={Journal of Shanghai University (English Edition)},
  volume={4},
  number={1},
  pages={22--26},
  year={2000},
  publisher={Springer}
}

@inproceedings{roman2024proactive,
author = {Roman, Robin San and Fernandez, Pierre and Elsahar, Hady and D\'{e}fossez, Alexandre and Furon, Teddy and Tran, Tuan},
title = {{Proactive detection of voice cloning with localized watermarking}},
year = {2024},
publisher = {JMLR.org},
booktitle = {Proceedings of the 41st International Conference on Machine Learning},
articleno = {1759},
numpages = {17},
location = {Vienna, Austria},
series = {ICML'24}
}

@misc{chen2023wavmark,
      title={{WavMark: Watermarking for Audio Generation}}, 
      author={Guangyu Chen and Yu Wu and Shujie Liu and Tao Liu and Xiaoyong Du and Furu Wei},
      year={2023},
      eprint={2308.12770},
      archivePrefix={arXiv},
      primaryClass={cs.SD}
}

@article{defossez2023high,
title={{High Fidelity Neural Audio Compression}},
author={Alexandre D{\'e}fossez and Jade Copet and Gabriel Synnaeve and Yossi Adi},
journal={Transactions on Machine Learning Research},
issn={2835-8856},
year={2023},
url={https://openreview.net/forum?id=ivCd8z8zR2},
note={Featured Certification, Reproducibility Certification}
}

@inproceedings{wang-etal-2021-voxpopuli,
    title = "{{VoxPopuli: A Large-Scale Multilingual Speech Corpus for Representation Learning, Semi-Supervised Learning and Interpretation}}",
    author = "Wang, Changhan  and
      Riviere, Morgane  and
      Lee, Ann  and
      Wu, Anne  and
      Talnikar, Chaitanya  and
      Haziza, Daniel  and
      Williamson, Mary  and
      Pino, Juan  and
      Dupoux, Emmanuel",
    editor = "Zong, Chengqing  and
      Xia, Fei  and
      Li, Wenjie  and
      Navigli, Roberto",
    booktitle = "Proceedings of the 59th Annual Meeting of the Association for Computational Linguistics and the 11th International Joint Conference on Natural Language Processing (Volume 1: Long Papers)",
    month = aug,
    year = "2021",
    address = "Online",
    publisher = "Association for Computational Linguistics",
    url = "https://aclanthology.org/2021.acl-long.80/",
    doi = "10.18653/v1/2021.acl-long.80",
    pages = "993--1003",
    
}

@article{WANG2026101825,
title = {{ASVspoof 5: Design, collection and validation of resources for spoofing, deepfake, and adversarial attack detection using crowdsourced speech}},
journal = {Computer Speech \& Language},
volume = {95},
pages = {101825},
year = {2026},
issn = {0885-2308},
doi = {https://doi.org/10.1016/j.csl.2025.101825},
url = {https://www.sciencedirect.com/science/article/pii/S0885230825000506},
author = {Xin Wang and Héctor Delgado and Hemlata Tak and Jee-weon Jung and Hye-jin Shim and Massimiliano Todisco and Ivan Kukanov and Xuechen Liu and Md Sahidullah and Tomi Kinnunen and Nicholas Evans and Kong Aik Lee and Junichi Yamagishi and Myeonghun Jeong and Ge Zhu and Yongyi Zang and You Zhang and Soumi Maiti and Florian Lux and Nicolas Müller and Wangyou Zhang and Chengzhe Sun and Shuwei Hou and Siwei Lyu and Sébastien {Le Maguer} and Cheng Gong and Hanjie Guo and Liping Chen and Vishwanath Singh},

}

\end{document}